\documentclass[aps,prd,onecolumn,showpacs,showkeys,amsmath,amssymb]{revtex4}
\usepackage{graphicx}
\usepackage{dcolumn}
\usepackage{bm}
\newcommand{\be}{\begin{eqnarray}}
\newcommand{\ee}{\end{eqnarray}}

\newcommand{\ket}{\rangle}
\newcommand{\bra}{\langle}
\def\lsim{\ \raise.3ex\hbox{$<$\kern-.75em\lower1ex\hbox{$\sim$}}\ }
\def\gsim{\ \raise.3ex\hbox{$>$\kern-.75em\lower1ex\hbox{$\sim$}}\ }
\begin{document}
\title{Exotic Tetraquark ud bar[s] bar[s] of JP=0+ in the QCD Sum
Rule}
%
\author{Hua-Xing Chen}
\email{hxchen@rcnp.osaka-u.ac.jp} \affiliation{Research Center for
Nuclear Physics, Osaka University, Ibaraki 567--0047, Japan and
Department of Physics, Peking University, Beijing 100871, China}
\author{Atsushi Hosaka}
\email{hosaka@rcnp.osaka-u.ac.jp} \affiliation{Research Center for
Nuclear Physics, Osaka University, Ibaraki 567--0047, Japan}
\author{Shi-Lin Zhu}
\email{zhusl@th.phy.pku.edu.cn} \affiliation{Department of Physics,
Peking University, Beijing 100871, China}
\begin{abstract}
We study a QCD sum rule analysis for an exotic tetraquark ud bar[s]
bar[s] of JP=0+ and I = 1. We construct q q bar[q] bar[q] currents
in a local product form and find that there are five independent
currents for this channel. Due to high dimensional nature of the
current, it is not easy to form a good sum rule when using a single
current. This means that we do not find any sum rule window to
extract reliable results, due to the insufficient convergence of the
OPE and to the exceptional important role of QCD continuum. Then we
examine sum rules by using currents of linear combinations of two
currents among the independent ones. We find two reasonable cases
that predict a mass of the tetraquark around 1.5 GeV.
\end{abstract}
\maketitle
\pagenumbering{arabic}
%
%
%
\section{Introduction}\label{sec_intro}
%

The history of exotic hadrons is rather long. But the recent
experimental observations have triggered tremendous amount of
research activities~\cite{Zhu:2004xa,Hicks:2004ge,Dzierba:2004db}.
Among them the report on the pentaquark $\Theta^+$ from the LEPS
group in 2002 was the most influential one~\cite{Nakano:2003qx},
partly because $\Theta^+$ is a genuine exotic state of the quark
content $uudd \bar s$. It also has unusual properties such as a
light mass and a very narrow width. Its existence is, however, now
questioned, which should be confirmed in the future
experiments~\cite{Schumacher:2005wu}.

Turning to mesons, though not genuine exotic states, $X(3872)$ and
$D_s(2317)$ are found to have properties that seem difficult to be
explained by a conventional picture of $\bar q
q$~\cite{Abe:2003hq,Choi:2003ue,Acosta:2003zx,Aubert:2003fg,Besson:2003cp,Krokovny:2003zq,Abe:2003jk}.
Rather, they could be  considered to have a significant amount of
multi-quark components. Historically, tetraquark mesons were
investigated long ago as an attempt to explain relatively light
masses and excess of states in scalar
channels~\cite{Jaffe:1976ig,Jaffe:1976ih,Jaffe:1976yi,Weinstein:1982gc,Chao:1980dv}.
Just as in the exotic baryons, it is interesting to consider genuine
exotic states in the meson sector whose minimal component is $qq
\bar q \bar q$. Tetraquark states of $ud \bar{s}\bar{s}$ component
have been studied as candidates of such exotic states. Since they
may be obtained by replacing one of $ud$ diquarks in $\Theta^+$ by
an $\bar{s}$ antiquark, similarities between $\Theta^+$ and $ud\bar
s\bar s$ have been discussed, though precise analogy is a dynamical
question~\cite{Zhu:2004za,Karliner:2004sy,Lichtenberg:2004tb}.

In the former studies, the tetraquark $ud\bar s \bar s$ of $J^P =
1^+$ was investigated in detail, where it was shown that the state
has a relatively low mass and a narrow width decaying into $K^* K$
in the flux tube model~\cite{Kanada-En'yo:2005zg}. The narrow decay
width is associated with the fact that $KK$ channel is forbidden due
to the conservation of parity and angular momentum, which partly
motivated the study of the $1^+$ channel.

In principle, it is also possible to study other channels of the $ud
\bar s \bar s$
tetraquarks~\cite{Burns:2004wy,Kanada-En'yo:2005zg,Cui:2005az}. From
a naive point of view of mass, it is natural to investigate $0^+$
scalar states. In contrast to $\bar qq$ mesons, the tetraquark does
not need orbital excitation to form the quantum number $0^+$, but
all quarks may occupy the lowest state. In this case, it is shown
that the tetraquark should have isospin one $I = 1$. This is the
object that we would like to study in this paper.

We perform QCD sum rule analyses for the scalar ($J^P = 0^+$) and
isovector ($I = 1$) exotic tetraquark $ud \bar s \bar s$. We attempt
a rather comprehensive analysis in which we will pay special
attention to the structure of the interpolating fields (currents).
First, we find that there are five independent interpolating fields
for the tetraquark. We show this by constructing the tetraquark
currents in terms of diquark fields ($(qq)(\bar q\bar q)$) and
mesonic fields ($(\bar qq)(\bar q q)$), where $\bar q q$ can be both
color singlet and octet. We then consider two-point correlation
functions first by using a single current of various types. It turns
out that many of them do not achieve a good sum rule. Therefore, we
attempt linear combinations of two independent currents. This method
was first proposed in Ref.~\cite{Wei:2004tc}. We then find that
there are several cases with good Borel stability, indicating the
mass of the tetraquark around $1.5$ GeV. We also investigate the
reliability of the sum rule not only from the Borel stability but
also from the dependence on the threshold value and the amount of
the pole contribution in the total sum rule. We also mention the
convergence of OPE.

The difficulties to make a good sum rule for exotic particles of
high dimensional operators were nicely discussed in a recent work by
Kojo et al.~\cite{Kojo:2006bh}. They proposed a sum rule using a
linear combination of two-point functions rather than currents in
order, for instance, to suppress large contributions from low
dimensional terms that are irrelevant to non-perturbative properties
of hadrons. They have successfully achieved a good sum rule that
satisfy the necessary requirements. In our present study, our
strategy is different from theirs, but the consideration along their
idea is certainly important in the discussion of the tetraquark
also.

This paper is organized as follows. In section 2, we establish five
independent currents in diquark-antidiquark and meson-meson
(actually meson-like) constructions. Some relations among various
currents will be discussed. Section 3 is the main part of this
paper, where we perform sum rule analyses using various tetraquark
currents constructed in section 2. We study the sum rule of a single
current and then consider linear combinations of currents. Section 4
is devoted to summary. In Appendix, we discuss the equivalence and
relations between the currents of diquark-antidiquark and
meson-meson constructions.

%
\section{Independent Currents}\label{sec_independent
current}
%

Let us consider currents for the tetraquark $ud\bar{s}\bar{s}$
having $J^P=0^+$. Here we consider only local currents. To write a
current, Lorentz and color indices are contracted with suitable
coefficients ($L^{abcd}_{\mu\nu\rho\sigma}$) to provide necessary
quantum numbers,
%
\begin{equation}
\label{define_eta} \eta = L^{abcd}_{\mu\nu\rho\sigma}\bar{s}_a^\mu
\bar{s}_b^\nu u_c^\rho d_d^\sigma\, ,
\end{equation}
%
where the sum over repeated indices ($\mu$, $\nu, \cdots$ for Dirac
spinor indices, and $a, b, \cdots$ for color indices) is taken.

For the Dirac spinor space, using possible diquark and antidiquark
bilinears~\cite{Jaffe:2003ci,Toki:2005vr,Karliner:2006hf,Selem:2006nd},
there are five independent terms
%
\begin{eqnarray}
\label{diquark_dirac_space}
\nonumber && S_{abcd} =
(\bar{s}_a\gamma_5 C \bar{s}_b^T)(u_c^T C \gamma_5 d_d) \, ,
\\ \nonumber &&
V_{abcd} = (\bar{s}_a \gamma_\mu \gamma_5 C \bar{s}_b^T)(u_c^T C
\gamma^\mu \gamma_5 d_d) \, ,
\\ &&
T_{abcd} = (\bar{s}_a \sigma_{\mu\nu} C \bar{s}_b^T)(u_c^T C \sigma^{\mu\nu} d_d) \, ,
\\ \nonumber &&
A_{abcd} = (\bar{s}_a \gamma_\mu C \bar{s}_b^T)(u_c^T C \gamma^\mu
d_d) \, ,
\\ \nonumber &&
P_{abcd} =(\bar{s}_a C \bar{s}_b^T)(u_c^T C d_d) \, .
\end{eqnarray}
%
Here, color indices are not yet specified. For the diquark and
antidiquark pair, color structures providing a color-singlet
tetraquark are $\mathbf{3}\otimes \mathbf{\bar 3}$ and $\mathbf{\bar
6} \otimes \mathbf{6}$, which we will denote by labels $\mathbf{3}$
and $\mathbf{6}$ for short.

Therefore, we have altogether ten terms of products
%
\begin{equation}
\label{diquark_product}
\{S \oplus V \oplus T \oplus A \oplus P\}_{Lorentz} \otimes
\{\mathbf{3} \oplus \mathbf{6}\}_{Color}
\, .
\end{equation}
%
However, half of them drop due to the Pauli principle. For instance
%
\begin{eqnarray}
\label{drop_P3}
P_3 &\equiv& P_{Lorentz} \otimes \mathbf{3}_{Color}
\\ \nonumber &=&
\epsilon_{abc} (\bar{s}_b C \bar{s}^T_c) \epsilon_{a b^\prime
c^\prime}(u^T_{b^\prime} C d_{c^\prime}) = 0 \, .
\end{eqnarray}
%
Eventually, we end up with five independent currents
%
\begin{eqnarray}
\nonumber\label{define_diquark_current}
&& S_6 = (\bar{s}_a \gamma_5 C \bar{s}_b^T)(u_a^T C \gamma_5 d_b)\,
,
\\ \nonumber &&
V_6 = (\bar{s}_a \gamma_\mu \gamma_5 C \bar{s}_b^T)(u_a^T C
\gamma^\mu \gamma_5 d_b)\, ,
\\ &&
T_3 = (\bar{s}_a \sigma_{\mu\nu} C \bar{s}_b^T)(u_a^T C
\sigma^{\mu\nu} d_b)\, ,
\\ \nonumber &&
A_3 = (\bar{s}_a \gamma_\mu C \bar{s}_b^T)(u_a^T C \gamma^\mu d_b)\,
,
\\ \nonumber &&
P_6 = (\bar{s}_a C \bar{s}_b^T)(u_a^T C d_b)\, .
\end{eqnarray}
%
In the non-relativistic language, these five terms correspond to
combinations of diquarks and antidiquarks
%
\begin{eqnarray}
\label{diquark_representation}
[(^1 S_0)(^1 S_0)]_{0^+}\, ,~~~[(^3 S_1)(^3 S_1)]_{0^+}\, ,~~~[(^1
P_1)(^1 P_1)]_{0^+}\, ,~~~[(^3 P_0)(^3 P_0)]_{0^+}\, ,~~~[(^3
P_1)(^3 P_1)]_{0^+}\, .
\end{eqnarray}
%
Another possible piece of $^3 P_2$ is irrelevant, since the five
bi-linear forms $q^T \Gamma q$ ($\Gamma = S, V, T, A, P$) can only
have spin $j \leq 1$, while the $^3 P_2$ diquark has $j=2$.

\begin{figure}[hbt]
\begin{center}
{\includegraphics[width=5cm]{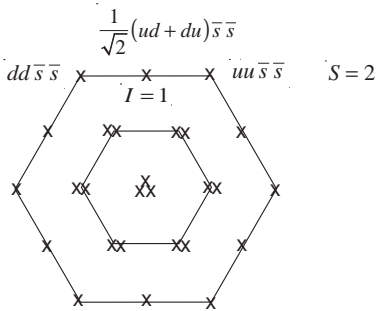}}
\caption{$SU(3)$ weight
diagram for $\mathbf{27}$, where the locations of three tetraquark
components of $S=2$ and $I=1$ are shown.}
\label{tetra27}
\end{center}
\end{figure}

Finally we consider the flavor structure. The $\bar{s}\bar{s}$
antidiquark is symmetric in flavor, and hence belongs to the
symmetric representation $\mathbf{\bar 6}_f$. If the other $ud$
diquark belongs to $\mathbf{\bar 3}_f$, and so isospin $I = 0$, the
diquark and antidiquark will have different flavor symmetry. But
they should have the same color and spin symmetries for composing a
color-singlet scalar tetraquark. Considering the Pauli principle,
they must have different parity, and hence their combination is a
negative-parity scalar tetraquark. Accordingly, the other $ud$
diquark also belongs to $\mathbf{6}_f$, and so isospin $I=1$. Among
the irreducible representations of the tetraquark
%
\begin{equation}
\nonumber
\mathbf{\bar{6}} \otimes \mathbf{6} = \mathbf{1} \oplus \mathbf{8}
\oplus \mathbf{27}\, ,
\end{equation}
%
$S=+2$ and $I=1$ states are in the $\mathbf{27}$ representation of
$SU(3)_f$, which is the flavor structure of the present tetraquark.
As shown in Fig.~\ref{tetra27}, three isovector states of the
$\mathbf{27}_f$ are $uu \bar s \bar s$, $1/\sqrt{2}(ud+du)\bar s\bar
s$ and $dd \bar s \bar s$.

%
%

We have constructed five independent currents using diquark and
antidiquark combination. We refer to this as the diquark
construction. Similarly, we can also construct the tetraquark
currents using $\bar{q}q$ combination (mesonic construction).
Obviously, there are ten combinations of the Dirac ($S$, $V$, $T$,
$A$ and $P$) and color ($\mathbf{1}$ and $\mathbf{8}$) spaces:
%
\begin{eqnarray}\nonumber\label{define_meson_currents}
&&S_1 = (\bar{s}_au_a)(\bar{s}_bd_b)\, ,~~~~~~~~~~~~~~~~~~~~~~~~~~~~
S_8 = (\bar{s}_a{\lambda^n_{ab}}u_b)(\bar{s}_c{\lambda^n_{cd}}d_d)\,
,
\\ \nonumber &&
V_1 = (\bar{s}_a\gamma_\mu u_a)(\bar{s}_b\gamma^\mu
d_b)\, ,~~~~~~~~~~~~~~~~~~~~~~
V_8 = (\bar{s}_a\gamma_\mu
{\lambda^n_{ab}} u_b)(\bar{s}_c\gamma^\mu {\lambda^n_{cd}} d_d)\, ,
\\ &&
T_1 = (\bar{s}_a\sigma_{\mu\nu}u_a)(\bar{s}_b\sigma^{\mu\nu}d_b)\,
,~~~~~~~~~~~~~~~~~~~
T_8 = (\bar{s}_a\sigma_{\mu\nu} {\lambda^n_{ab}}
u_b)(\bar{s}_c\sigma^{\mu\nu} {\lambda^n_{cd}} d_d)\, ,
\\ \nonumber &&
A_1 = (\bar{s}_a\gamma_\mu\gamma_5u_a)(\bar{s}_b\gamma^\mu\gamma_5d_b
)\, ,~~~~~~~~~~~~~~~
A_8 = (\bar{s}_a\gamma_\mu\gamma_5
{\lambda^n_{ab}} u_b)(\bar{s}_c\gamma^\mu\gamma_5 {\lambda^n_{cd}}
d_d)\, ,
\\ \nonumber &&
P_1 = (\bar{s}_a\gamma_5u_a)(\bar{s}_b\gamma_5d_b)\,
,~~~~~~~~~~~~~~~~~~~~~~
P_8 = (\bar{s}_a\gamma_5 {\lambda^n_{ab}}
u_b)(\bar{s}_c\gamma_5 {\lambda^n_{cd}} d_d)\, ,
\end{eqnarray}
%
where subscripts $\mathbf{1}$ and $\mathbf{8}$ denote
color singlet and octet
representations, respectively.
Unlike the diquark construction, all the ten
currents in Eq.~(\ref{define_meson_currents}) remain finite.
However, it is possible to show only five of them (in fact any five
of them) are independent. The proof of this and various relations
among different currents are discussed in
Appendix.~\ref{app_five_currents}.

%
\section{QCD Sum Rules Analysis}\label{sec_sum_rule}
%
%
\subsection{Formulae of QCD Sum Rule}\label{subsec_formalism}
%

For the past decades QCD sum rule has proven to be a very powerful
and successful non-perturbative
method~\cite{Shifman:1978bx,Reinders:1984sr}. In sum rule analyses,
we consider two-point correlation functions:
%
\begin{equation}
\Pi(q^2)\,\equiv\,i\int d^4x e^{iqx}
\langle0|T\eta(x){\eta^\dagger}(0)|0\rangle \, ,
\label{eq_pidefine}
\end{equation}
%
where $\eta$ is an interpolating current for the tetraquark. We
compute $\Pi(q^2)$ in the operator product expansion (OPE) of QCD up
to certain order in the expansion, which is then matched with a
hadronic parametrization to extract information of hadron
properties. At the hadron level, we express the correlation function
in the form of the dispersion relation with a spectral function:
%
\begin{equation}
\Pi(p)=\int^\infty_0\frac{\rho(s)}{s-p^2-i\varepsilon}ds \, ,
\label{eq_disper}
\end{equation}
%
where
%
\begin{eqnarray}
\rho(s) & \equiv & \sum_n\delta(s-M^2_n)\langle
0|\eta|n\rangle\langle n|{\eta^\dagger}|0\rangle \ \nonumber\\ &=&
f^2_X\delta(s-M^2_X)+ \rm{higher\,\,states}\, . \label{eq_rho}
\end{eqnarray}
%
For the second equation, as usual, we adopt a parametrization of one
pole dominance for the ground state $X$ and a continuum
contribution. The sum rule analysis is then performed after the
Borel transformation of the two expressions of the correlation
function, (\ref{eq_pidefine}) and (\ref{eq_disper})
%
\begin{equation}
\Pi^{(all)}(M_B^2)\equiv\mathcal{B}_{M_B^2}\Pi(p^2)=\int^\infty_0
e^{-s/M_B^2} \rho(s)ds \, . \label{eq_borel}
\end{equation}
%
Assuming the contribution from the continuum states can be
approximated well by the spectral density of OPE above a threshold
value $s_0$ (duality), we arrive at the sum rule equation
%
\begin{equation}
\Pi(M_B^2) \equiv f^2_Xe^{-M_X^2/M_B^2} = \int^{s_0}_0
e^{-s/M_B^2}\rho(s)ds \label{eq_fin} \, .
\end{equation}
%
Differentiating Eq.~(\ref{eq_fin}) with respect to {\Large
$\frac{1}{M_B^2}$} and dividing it by Eq. (\ref{eq_fin}), finally we
obtain
%
\begin{equation}
M^2_X=\frac{\int^{s_0}_0
e^{-s/M_B^2}s\rho(s)ds}{\int^{s_0}_0 e^{-s/M_B^2}\rho(s)ds}\, .
\label{eq_LSR}
\end{equation}
%
In the following, we study both Eqs.~(\ref{eq_fin}) and
(\ref{eq_LSR}) as functions of the parameters such as the Borel mass
$M_B$ and the threshold value $s_0$ for various combinations of the
tetraquark currents.

%
\subsection{Analysis of Single Diquark
Currents}\label{sec_diquark_single_current}
%

In this subsection, we perform a QCD sum rule analysis using the
five diquark currents, $S_6$, $V_6$, $T_3$, $A_3$ and $P_6$,
separately. Let us first outline briefly how we performed the OPE
calculation. For illustration, let us take $P_6$. Then
%
\be \Pi(q^2)\, &\equiv&\, i\int d^4x e^{iqx} \langle0|T
P_6(x){P_6^\dagger}(0)|0\rangle
\nonumber \\
&=& Tr[C (S_u^{a a^\prime}(x))^T C S_d^{b b^\prime}(x)]
Tr[S_s^{a^\prime a}(-x) C (S_s^{b^\prime b}(-x))^T C]
\\ \nonumber &&+Tr[C (S_u^{a a^\prime}(x))^T C
S_d^{b b^\prime}(x)]  Tr[S_s^{b^\prime a}(-x) C (S_s^{a^\prime
b}(-x))^T C]\, . \ee
%
For the quark propagator, we use
%
\begin{eqnarray}\nonumber\label{app_eq_propagator}
i\mbox{S}^{ab}_q(x)&\equiv&\langle0|\mbox{T}[q^a(x)\bar{q}^b(0)]|0\rangle
\\ &=&\frac{i\delta^{ab}}{2\pi^2x^4}\hat{x}
+\frac{i}{32\pi^2}\frac{\lambda^n_{ab}}{2}\mbox{g}_c\mbox{G}^n_{\mu\nu}\frac{1}{x^2}(\sigma^{\mu\nu}\hat{x}+\hat{x}\sigma^{\mu\nu})
-\frac{\delta^{ab}}{12}\langle\bar{q}q\rangle
+\frac{\delta^{ab}x^2}{192}\langle g_c\bar{q}\sigma Gq\rangle
\\ \nonumber &&-\frac{\delta^{ab}m_q}{4\pi^2x^2}
+\frac{i\delta^{ab}m_q }{48}\langle\bar{q}q\rangle\hat{x}
+\frac{i\delta^{ab}m_q^2}{8\pi^2x^2}\hat{x}\, .
\end{eqnarray}
%
The two-point function is then divided into three parts:
%
\begin{enumerate}

\item
Terms proportional to $\delta^{ab}$ ($a, b$ being color indices),
where no soft gluon is emitted.
The lowest term of this kind is the continuum term.

\item
Terms containing one $\lambda_{ab}$ (color matrix), where
one soft gluon is emitted.
The lowest terms of this type contain
condensates such as
$\langle g \bar{q} \sigma G q \rangle$
($q = u$ and $d$) and $\langle
g \bar{s} \sigma G s \rangle$.

\item
Terms containing two $\lambda_{ab}$'s, where
two soft gluons are emitted.
The lowest terms of this type contain the condensate
$\langle g^2 G^2 \rangle$.
\end{enumerate}
%

We have performed the OPE calculation for the spectral function up
to dimension eight, which is up to the constant ($s^0$) term of
$\rho(s)$. Actual computation is very complicated. We have performed
this calculation using $Mathematica$ with
$FeynCalc$~\cite{feyncalc}. $Mathematica$ programs are available
from the authors. The results are
\be \rho_{S6}(s)&=&\frac{s^4} {61440 \pi^6} -\frac{{m_s}^2 s^3}
{3072 \pi^6} + \Big ( \frac{{m_s}^4} {256 \pi^6}  -\frac{{m_s}
\langle \bar{s}s \rangle} {192 \pi^4} -\frac{\langle g^2GG \rangle}
{12288 \pi^6} \Big  ) s^2
\nonumber \\
&& + \Big  ( -\frac{m_s^3 \langle \bar{s}s \rangle}{32\pi^4} +
\frac{m_s^2 \langle g^2GG \rangle}{4096\pi^6} - \frac{m_s \langle
g\bar{s} \sigma Gs \rangle}{64\pi^4} + \frac{\langle \bar{q}q
\rangle ^2}{24\pi^2} + \frac{\langle \bar{s}s \rangle ^2}{24\pi^2}
\Big ) s
\\
\nonumber && - \frac{m_s^2 \langle \bar{q}q \rangle ^2}{12\pi^2} +
\frac{m_s^2 \langle \bar{s}s \rangle^2} {48\pi^2} + \frac{\langle
\bar{q}q \rangle \langle g\bar{q}\sigma Gq \rangle}{24\pi^2} +
\frac{m_s\langle g^2 GG \rangle\langle \bar{s}s \rangle}{1536\pi^4}
+ \frac{\langle \bar{s}s \rangle\langle g\bar{s} \sigma Gs
\rangle}{24\pi^2} - \frac{m_s^4\langle g^2 GG \rangle}{2048\pi^6}\,
,
\\
\nonumber \rho_{V6}(s)&=&\frac{s^4} {15360 \pi^6} -\frac{5{m_s}^2
s^3} {1536 \pi^6} + \Big (\frac{{m_s}^4} {64 \pi^6} + \frac{{m_s}
\langle \bar{s}s \rangle} {24 \pi^4} + \frac{5\langle g^2GG \rangle}
{6144 \pi^6}  \Big ) s^2
\\
&& +  \Big ( -\frac{m_s^3 \langle \bar{s}s \rangle}{8\pi^4} -
\frac{11 m_s^2 \langle g^2GG \rangle}{2048\pi^6} + \frac{m_s \langle
g\bar{s} \sigma Gs \rangle}{32\pi^4} -\frac{\langle \bar{q}q \rangle
^2}{12\pi^2} -\frac{\langle \bar{s}s \rangle ^2}{12\pi^2} \Big ) s
\\ \nonumber &&
+\frac{2m_s^2 \langle \bar{q}q \rangle ^2}{3\pi^2}  +\frac{m_s^2
\langle \bar{s}s \rangle^2} {12\pi^2} -\frac{\langle \bar{q}q
\rangle \langle g\bar{q}\sigma Gq \rangle}{12\pi^2} + \frac{7 m_s
\langle g^2GG \rangle \langle \bar{s}s \rangle}{768\pi^4} -
\frac{\langle \bar{s}s \rangle\langle g\bar{s} \sigma Gs
\rangle}{12\pi^2}\, ,
\\
\nonumber \rho_{T3}(s)&=&\frac{s^4} {5120 \pi^6} -\frac{{m_s}^2 s^3}
{128 \pi^6} + \Big (\frac{3{m_s}^4} {64 \pi^6}  +\frac{{m_s} \langle
\bar{s}s \rangle} {16 \pi^4}  + \frac{\langle g^2GG \rangle} {1536
\pi^6}  \Big ) s^2 +  \Big ( -\frac{3 m_s^3 \langle \bar{s}s
\rangle}{8\pi^4} - \frac{m_s^2 \langle g^2GG \rangle}{256\pi^6} \Big
) s
\\
&& +\frac{m_s^2 \langle \bar{q}q \rangle ^2}{\pi^2}  + \frac{m_s^2
\langle \bar{s}s \rangle^2} {4\pi^2} + \frac{m_s\langle g^2 GG
\rangle\langle \bar{s}s \rangle}{192\pi^4} - \frac{m_s^4\langle g^2
GG \rangle}{256\pi^6}\, ,
\\
\nonumber \rho_{A3}(s)&=&\frac{s^4} {30720 \pi^6} -\frac{{m_s}^2
s^3} {1024 \pi^6} +  \Big ( \frac{{m_s}^4} {128 \pi^6} +
\frac{\langle g^2GG \rangle} {6144 \pi^6}  \Big ) s^2
\\
&& +  \Big ( -\frac{m_s^3 \langle \bar{s}s \rangle}{16\pi^4} -
\frac{3 m_s^2 \langle g^2GG \rangle}{2048\pi^6} - \frac{m_s \langle
g\bar{s} \sigma Gs \rangle}{64\pi^4} + \frac{\langle \bar{q}q
\rangle ^2}{24\pi^2} + \frac{\langle \bar{s}s \rangle ^2}{24\pi^2}
\Big ) s
\\
\nonumber && + \frac{m_s^2 \langle \bar{s}s \rangle^2} {24\pi^2} +
\frac{\langle \bar{q}q \rangle \langle g\bar{q}\sigma Gq
\rangle}{24\pi^2} + \frac{m_s\langle g^2 GG \rangle\langle \bar{s}s
\rangle}{256\pi^4} + \frac{\langle \bar{s}s \rangle\langle g\bar{s}
\sigma Gs \rangle}{24\pi^2}\, ,
\\
\nonumber\label{rho_p6} \rho_{P6}(s)&=&\frac{s^4} {61440 \pi^6}
-\frac{{m_s}^2 s^3} {1024 \pi^6} + \Big (\frac{{m_s}^4} {256 \pi^6}
-\frac{{m_s} \langle \bar{s}s \rangle} {64 \pi^4}  -\frac{\langle
g^2GG \rangle} {12288 \pi^6} \Big  ) s^2
\\ && + \Big  (
-\frac{m_s^3 \langle \bar{s}s \rangle}{32\pi^4} +\frac{3 m_s^2
\langle g^2GG \rangle}{4096\pi^6} +\frac{m_s \langle g\bar{s} \sigma
Gs \rangle}{64\pi^4} -\frac{\langle \bar{q}q \rangle ^2}{24\pi^2}
-\frac{\langle \bar{s}s \rangle ^2}{24\pi^2} \Big ) s
\\ \nonumber &&
+\frac{m_s^2 \langle \bar{q}q \rangle ^2}{4\pi^2}  +\frac{m_s^2
\langle \bar{s}s \rangle^2} {48\pi^2} -\frac{\langle \bar{q}q
\rangle \langle g\bar{q}\sigma Gq \rangle}{24\pi^2} -
\frac{m_s\langle g^2 GG \rangle\langle \bar{s}s \rangle}{512\pi^4} -
\frac{\langle \bar{s}s \rangle\langle g\bar{s} \sigma Gs
\rangle}{24\pi^2} -\frac{m_s^4\langle g^2 GG \rangle}{2048\pi^6}\, .
\ee
%
In these equations, $q$ represents a $u$ or $d$ quark, and $s$
represents an $s$ quark. $\langle \bar{q}q \rangle$ and $\langle
\bar{s}s \rangle$ are dimension $D=3$ quark condensates; $\langle
g^2 GG \rangle$ is a $D=4$ gluon condensate; $\langle g\bar{q}\sigma
Gq \rangle$ and $\langle g\bar{s}\sigma Gs \rangle$ are $D=5$ mixed
condensates. From these expressions, we observe the followings:
%
\begin{itemize}

\item
The coefficients of the lowest dimension, or of the leading term in
powers of $s$, have the relations $c_{S6}^{(4)} = c_{P3}^{(4)}$ and
$c_{A3}^{(4)} = 1/2 c_{V6}^{(4)}$. These are the consequences of
chiral symmetry at the perturbative level~\cite{Hosaka:2001ux}.

\item
As empirically known,
the terms of quark condensates have important
contributions to the sum rule.

\end{itemize}
%

For numerical calculations, we use the following values of
condensates\cite{Yang:1993bp,Narison:2002pw,Gimenez:2005nt,Jamin:2002ev,Ioffe:2002be,Ovchinnikov:1988gk,Eidelman:2004wy}:
%
\begin{eqnarray}
\nonumber &&\langle\bar qq \rangle=-(0.240 \mbox{ GeV})^3\, ,
\\
\nonumber &&\langle\bar ss\rangle=-(0.8\pm 0.1)\times(0.240 \mbox{
GeV})^3\, ,
\\
\nonumber &&\langle g_s^2GG\rangle =(0.48\pm 0.14) \mbox{ GeV}^4\, ,
\\
\label{condensates} &&m_s(2\mbox{ GeV})=0.11 \mbox{ GeV}\, ,
\\
\nonumber && \langle g_s\bar q\sigma G
q\rangle=-M_0^2\times\langle\bar qq\rangle\, ,
\\
\nonumber &&M_0^2=(0.8\pm0.2)\mbox{ GeV}^2\, .
\end{eqnarray}
%
As usual we assume the vacuum saturation for higher dimensional
operators such as $\bra 0 | \bar q q \bar q q |0\ket \sim \bra 0 |
\bar q q |0\ket \bra 0|\bar q q |0\ket$. There is a minus sign in
the definition of the mixed condensate $\langle g_s\bar q\sigma G
q\rangle$, which is different with some other QCD sum rule
calculation. This is just because the definition of coupling
constant $g_s$ is different~\cite{Yang:1993bp,Hwang:1994vp}.

%
\begin{figure}[hbt]
\begin{center}
\scalebox{0.9}{\includegraphics{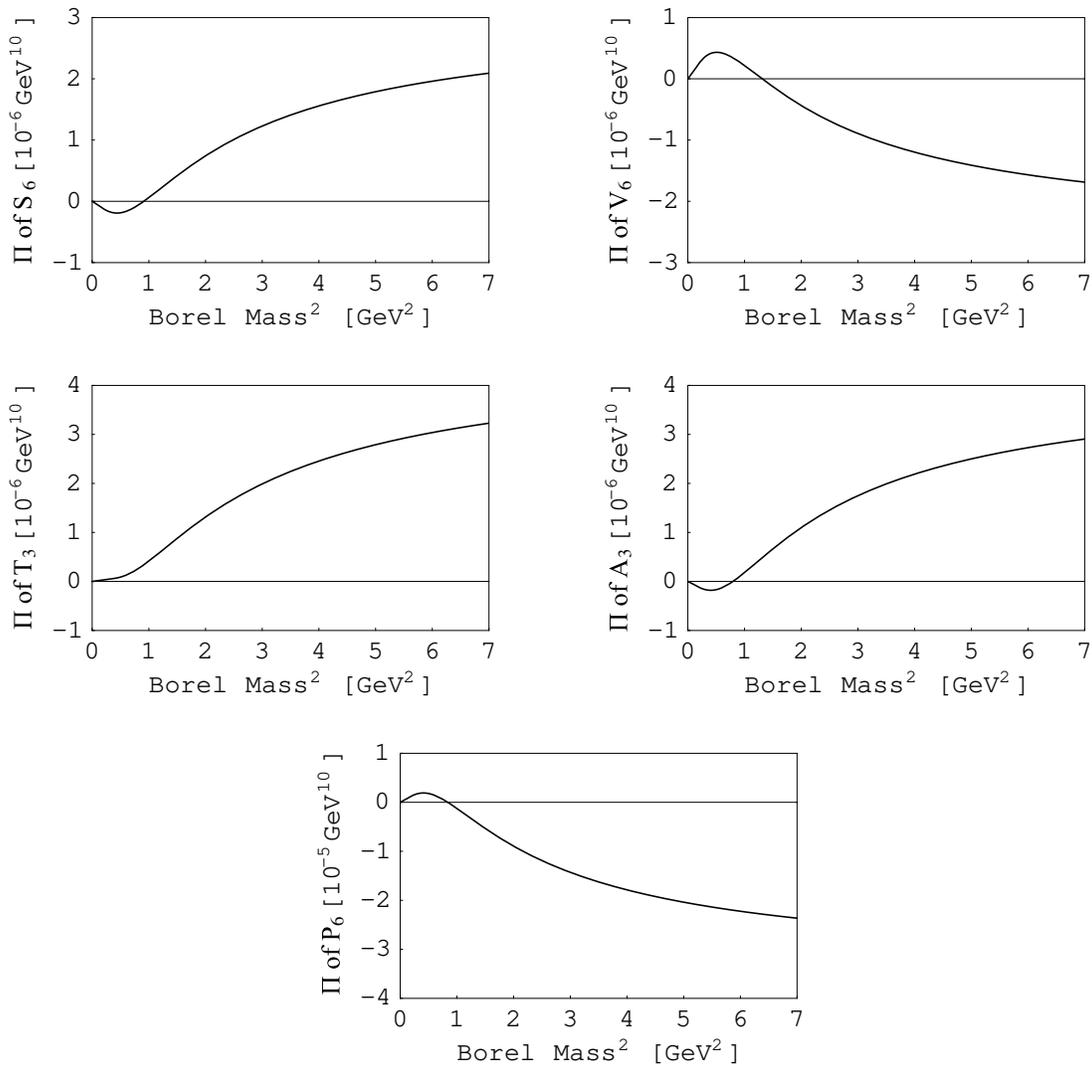}} \caption{Borel
transformed correlation functions $\Pi_{S6}$, $\Pi_{V6}$,
$\Pi_{T3}$, $\Pi_{A3}$ and $\Pi_{P6}$ as functions of Borel mass
square, in units of $\mbox{GeV}^{10}$, for threshold value $s_0 = 3$
GeV$^2$.} \label{pic_pi_diquark}
\end{center}
\end{figure}
%

In Fig.~\ref{pic_pi_diquark}, we show all five Borel transformed
correlation functions $\Pi(M_B^2)$ (the LHS of the
Eq.~(\ref{eq_fin})) as functions of Borel mass square for threshold
value $s_0 = 3$ GeV$^2$. From the definition of (\ref{eq_rho}), the
LHS should be positive definite quantities. In practical
calculations, however, the positivity may not be necessarily
realized, if the OPE up to finite terms does not work due to
insufficient convergence of the OPE. In the present analysis, we
find that among the five cases, two functions of $V_6$ and $P_6$
currents show such a bad behavior. Therefore, the QCD sum rules for
these two currents are not physically acceptable. The correlation
functions of $A_3$ and $S_6$ change the sign from negative to
positive values. But the sum rule values take positive values for
$M_B^2 \sim$ several GeV$^2$.

The tetraquark currents $S_6$ and $A_3$ are constructed by diquark
fields which correspond to $^1S_0$ and $^3S_1$ in the
non-relativistic language, where the two quarks can be in the ground
state $s$-orbit. In contrast, the currents $V_6$ and $P_6$
correspond to linear combinations of $^3P_1$, and $^3P_0$,
respectively, where one of the two quarks is in an excited
$p$-orbit. The $T_3$ current is a linear combination of $^3S_1$ and
$^1P_1$. Therefore, we verify an empirical fact that the sum rule
constructed by currents having the $s$-wave components in the
non-relativistic limit works better than those dominated by $p$-wave
components. For completeness, we show the LHS with numerical
coefficients for the three better cases: $A_3$, $T_3$ and $S_6$
%
\begin{eqnarray}\nonumber\label{rho_pol_qq}
\Pi_{A3}^{(all)} &=& 8.2 \times 10^{-7} M_B^{10} - 7.4 \times
10^{-8} M_B^8 + 1.6 \times 10^{-7} M_B^6 + 1.8 \times 10^{-6} M_B^4
- 1.1 \times 10^{-6} M_B^2\, ,
\\ \nonumber \Pi_{T3}^{(all)} &=& 4.8 \times 10^{-6} M_B^{10} - 5.9 \times 10^{-7}
M_B^8 - 9.1 \times 10^{-7} M_B^6 + 3.4 \times 10^{-8} M_B^4 + 2.4
\times 10^{-7} M_B^2\, ,
\\ \nonumber \Pi_{S6}^{(all)} &=& 4.1 \times 10^{-7} M_B^{10} - 2.5 \times 10^{-8}
M_B^8 + 5.1 \times 10^{-8} M_B^6 + 1.8 \times 10^{-6} M_B^4 - 1.1
\times 10^{-6} M_B^2\, .
\\
\end{eqnarray}
%
From these expressions, we observe that the convergence of the
current $T_3$ seems better, while the convergence of the currents
$A_3$ and $S_6$ is not very good in the region $1 < M_B^2 < 2$
GeV$^2$. They can only converge at $M_B^2 \sim 3$ GeV$^2$.

\begin{figure}[!hbp]
\begin{center}
\scalebox{0.9}{\includegraphics{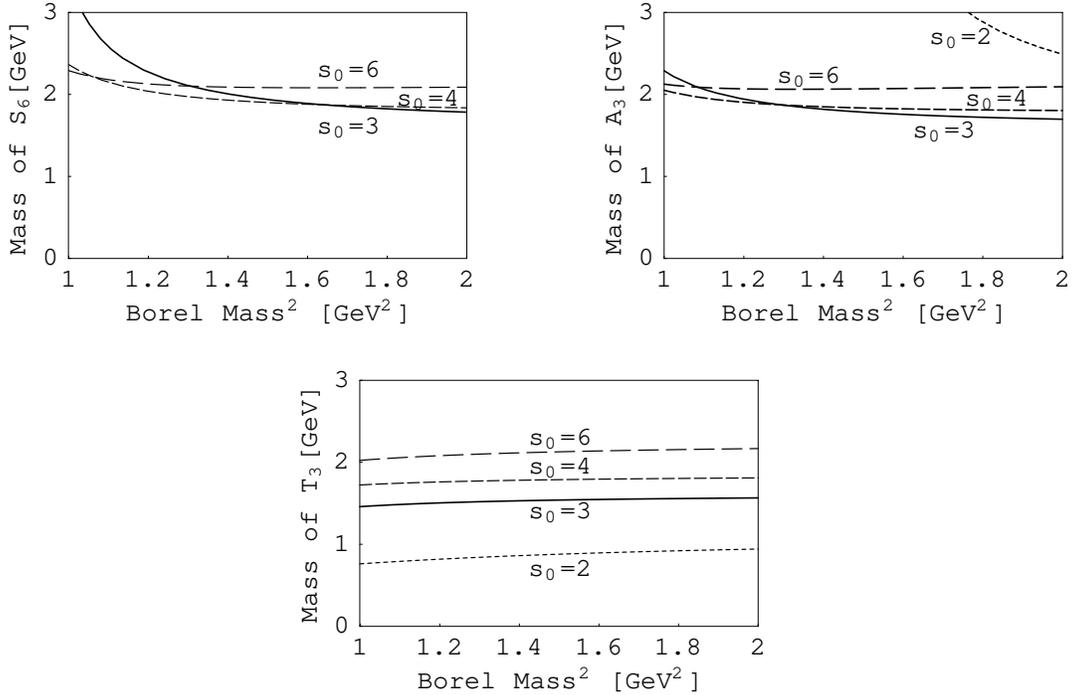}} \caption{Mass of
the tetraquark calculated by the three currents $S_6$, $A_3$ and
$T_3$, as functions of the Borel mass square $M_B^2$, for several
threshold values $s_0 = 2, 3, 4$ and 6 GeV$^2$.}
\label{pic_mass_diquark_single}
\end{center}
\end{figure}

To determine the mass, we need to fix the two parameters: the
threshold value $s_0$ and the Borel mass square $M_B^2$. For a good
sum rule, the predicted masses should not depend on these two
parameters strongly with sizable pole contribution (Borel window).
In Fig.~\ref{pic_mass_diquark_single}, we show the masses of the
tetraquark as functions of the Borel mass for several threshold
values $s_0$ (Borel curves). We observe that the Borel mass
dependence is somewhat strong for the currents $S_6$ and $A_3$ in
the region $1 < M_B^2 < 2$ GeV$^2$, which is expected to be a
reasonable choice of the Borel mass. For these currents $S_6$ and
$A_3$, however, we see that the minimum occurs at around 3 GeV$^2$
when $s_0$ is varied in the region $M_B^2 \gsim 1.5$ GeV$^2$. (For
the current $S_6$, the mass of $s_0 = 2$ GeV$^2$ is far above the
region shown in the figure.) For this reason, we consider that $s_0
=3$ GeV$^2$ is a reasonable choice which we will mainly use for the
estimation of the mass of the tetraquark in the following sum rule
analyses. At this $s_0$ value, the mass of the tetraquark turns out
to be about 1.6 GeV. For the $T_3$ current, the Borel stability
seems better. The result, however, depends on the threshold value
$s_0$ to some extent. However, it is interesting to see that the
mass of the tetraquark is about 1.6 GeV when $s_0 \sim 3$ GeV$^2$.

%
%

To see the amount of the pole contribution, we define the quantity
\begin{equation}\label{eq_pole}
\mbox{Pole contribution} \equiv \frac{\int_{4m_s^2}^{s_0}
e^{-s/M_B^2}\rho(s)ds}{\int_{4m_s^2}^{\infty}
e^{-s/M_B^2}\rho(s)ds}\, .
\end{equation}
As shown in Table 1, the pole contribution of the diquark currents
$A_3$, $T_3$ and $S_6$ are not very large; at $M_B^2 \sim 1$ GeV$^2$
they are of order 10 \%. This is a general problem of the QCD sum
rule when multi-quark currents are used. Therefore the results so
far might be doubtful.

\begin{table}
\caption{Pole contributions of various currents. The threshold value
$s_0 = 3$ GeV$^2$ is used.}
\begin{center}
\begin{tabular}{c|ccc|cc|cc}
\hline & \multicolumn{3}{c|}{Diquark Current} &
\multicolumn{2}{c|}{Mesonic Current}& \multicolumn{2}{c}{Mixed
Current}
\\  $M_B^2$ & $A_3$ & $T_3$ & $S_6$ & $V_8$ & $T_8$ & $\eta_1$
& $\eta_2$
\\ \hline 0.7 GeV$^2$ & --- & --- & --- & --- & --- & 0.60 & 0.49
\\ 1 GeV$^2$ & 0.17 & 0.11 & 0.10 & 0.54 & 0.23 & 0.30 & 0.22
\\ 2 GeV$^2$ & 0.04 & 0.01 & 0.05 & 0.09 & 0.02 & 0.03 & 0.02
\\ \hline
\end{tabular}
\end{center}
\end{table}

From the analysis of the single current of the diquark construction,
we expect that the mass of the tetraquark is about 1.6 GeV, although
the stability against the variation of both the Borel mass and the
threshold parameter is not simultaneously achieved. Furthermore, the
pole contribution is rather small. As we will see, however, a
suitable linear combination will improve them.

%
\subsection{Analysis of Single Mesonic
Currents}\label{sec_mesonic_single_current}
%

\begin{figure}[!hbp]
\begin{center}
\scalebox{0.75}{\includegraphics{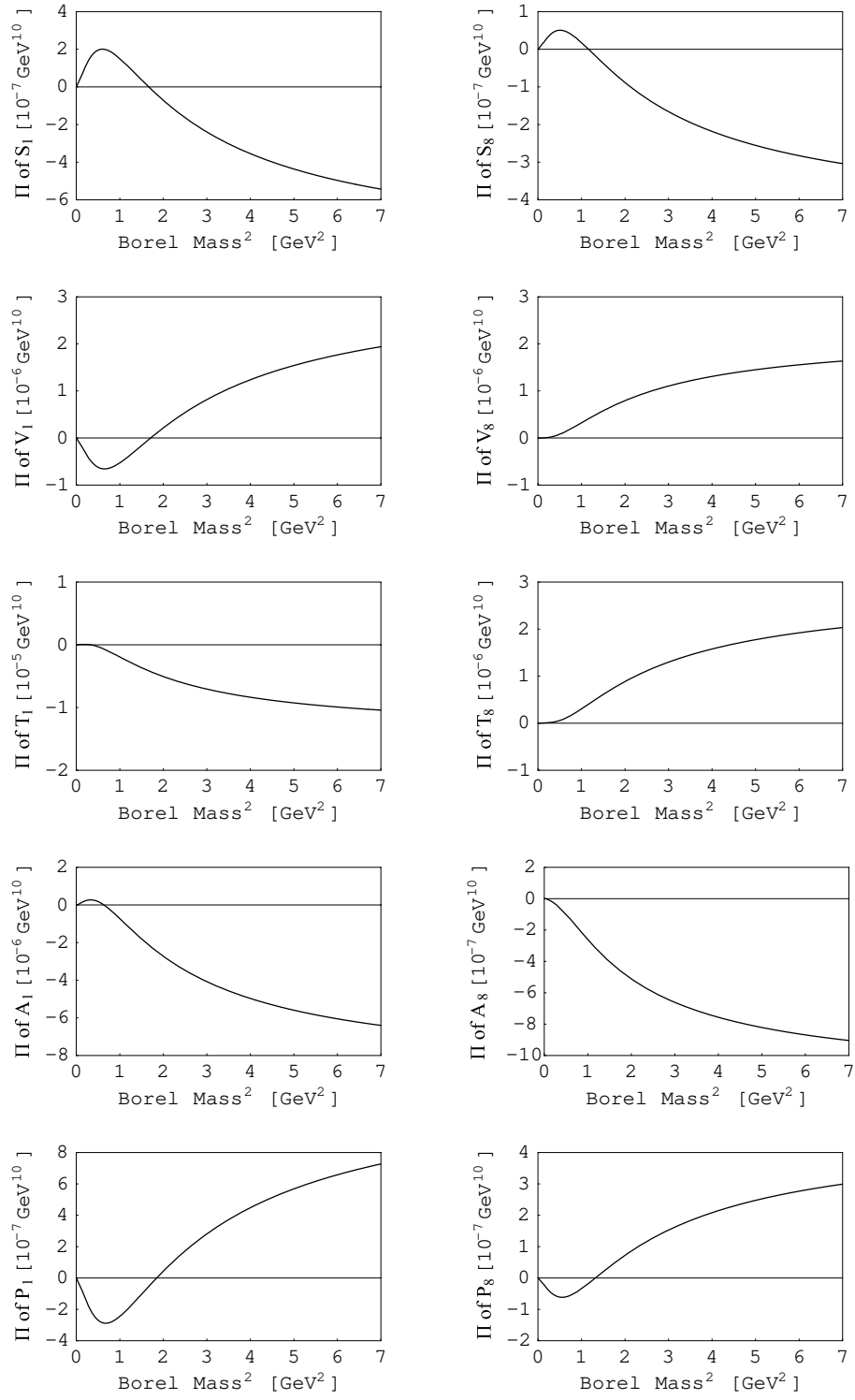}} \caption{Borel
transformed correlation functions $\Pi_{S1}$, $\Pi_{S8}$,
$\Pi_{V1}$, $\Pi_{V8}$, $\Pi_{T1}$, $\Pi_{T8}$, $\Pi_{A1}$,
$\Pi_{A8}$, $\Pi_{P1}$ and $\Pi_{P8}$ as functions of Borel mass
square, in units of $\mbox{GeV}^{10}$, for threshold value $s_0 = 3$
GeV$^2$.} \label{pic_rho_mesonic_single}
\end{center}
\end{figure}

In this subsection, we perform QCD sum rule analysis using the ten
mesonic currents, $S_{1,8}$, $V_{1,8}$, $T_{1,8}$, $A_{1,8}$ and
$P_{1,8}$, separately. Here we only show two important spectral
densities:
%
\begin{eqnarray} \nonumber \label{rho_v8}
\rho_{V8}(s)&=&\frac{s^4}{110592 \pi^6}  -\frac{19 {m_s}^2 s^3}
{55296\pi^6} +  \Big (\frac{5 {m_s}^4}{2304 \pi^6} - \frac{{m_s}
\langle \bar{q}q \rangle}{432 \pi^4} + \frac{{m_s}\langle \bar{s}s
\rangle} {432\pi^4} + \frac{17 \langle g^2GG \rangle}{221184 \pi^6}
\Big ) s^2
\\ \nonumber &&
+  \Big (\frac{m_s^3 \langle \bar{q}q \rangle}{72\pi^4} -
\frac{5m_s^3 \langle \bar{s}s \rangle}{288\pi^4} - \frac{13m_s^2
\langle g^2GG \rangle}{24576\pi^6} + \frac{m_s \langle g\bar{q}Gq
\rangle}{2304\pi^4} - \frac{5m_s \langle g\bar{s}Gs
\rangle}{4608\pi^4}
\\ &&
+ \frac{\langle \bar{q}q\rangle^2}{432\pi^2} + \frac{\langle
\bar{s}s \rangle ^2}{432\pi^2} + \frac{\langle \bar{q}q \rangle
\langle \bar{s}s \rangle}{54\pi^2} \Big ) s + \frac{m_s^2 \langle
\bar{q}q \rangle ^2}{27\pi^2} + \frac{5m_s^2 \langle \bar{s}s
\rangle^2}{432\pi^2}
\\ \nonumber &&
- \frac{m_s\langle \bar{q}q \rangle \langle g^2 GG
\rangle}{6912\pi^4} + \frac{5 \langle \bar{q}q \rangle \langle
g\bar{q}G q \rangle}{1728\pi^2} + \frac{m_s^3 \langle g\bar{q}Gq
\rangle}{144\pi^4}  - \frac{m_s^2 \langle \bar{q}q \rangle \langle
\bar{s}s \rangle}{18\pi^2} - \frac{\langle g\bar{q}Gq \rangle\langle
\bar{s}s \rangle}{864\pi^2}
\\ \nonumber &&
+ \frac{m_s\langle g^2 GG \rangle\langle \bar{s}s
\rangle}{1024\pi^4} - \frac{\langle \bar{q}q \rangle\langle
g\bar{s}Gs \rangle}{864\pi^2} + \frac{5\langle \bar{s}s
\rangle\langle g\bar{s}Gs \rangle}{1728\pi^2} - \frac{m_s^4\langle
g^2 GG \rangle}{9216\pi^6}\, ,
\\
\nonumber \label{rho_t8} \rho_{T8}(s) & = & \frac{s^4}{18432 \pi^6}
- \frac{5 {m_s}^2 s^3}{2304\pi^6} +  \Big ( \frac{5 {m_s}^4}{384
\pi^6} + \frac{5 {m_s} \langle \bar{s}s \rangle}{288 \pi^4} +
\frac{31 \langle g^2GG \rangle}{55296 \pi^6} \Big ) s^2
\\ &&
+  \Big ( -\frac{5 m_s^3 \langle \bar{s}s \rangle}{48\pi^4} -
\frac{31m_s^2 \langle g^2GG \rangle}{9216\pi^6}  \Big ) s +
\frac{5m_s^2 \langle \bar{q}q \rangle ^2}{18\pi^2}+ \frac{5m_s^2
\langle \bar{s}s \rangle^2}{72\pi^2} \\ \nonumber && +
\frac{31m_s\langle g^2 GG \rangle\langle \bar{s}s
\rangle}{6912\pi^4} - \frac{13m_s^4\langle g^2 GG
\rangle}{9216\pi^6}\, .
\end{eqnarray}
%
As shown in Fig.~\ref{pic_rho_mesonic_single}, we find that among
the ten correlation functions, only two correlation functions for
the currents $V_8$ and $T_8$ show good behavior with having positive
values.

%
%

The currents $V_1$, $V_8$, $P_1$ and $P_8$ are constructed by
mesonic fields (either color singlet or color octet) which
correspond to $^3S_1$ and $^1S_0$ in the non-relativistic language,
where two quark-antiquark pairs can be in the ground state
$s$-orbit. Their spectral densities then show similar behavior to
$S_6$ and $A_3$ in the previous subsection. In contrast, $S_1$,
$S_8$, $A_1$ and $A_8$ correspond to linear combinations of $^3P_0$
and $^3P_1$, respectively; $T_1$ and $T_8$ currents are the
combinations of $^3S_1$ and $^1P_1$.

From the above argument, we might expect that six currents, $V_1$,
$V_8$, $P_1$, $P_8$, $T_1$ and $T_8$ would work. However, we found
that the Borel transformed correlation functions calculated by the
currents $V_1$, $P_1$, $P_8$ and $T_1$ take negative values and
therefore, they must be abandoned. Now there remain only two better
currents $V_8$ and $T_8$ in the mesonic construction. This is the
reason why we have shown their spectral densities in (\ref{rho_v8})
and (\ref{rho_t8}). Using the numerical values of various
condensates (\ref{condensates}), we find the Borel transformed
correlation functions
%
\begin{eqnarray}\nonumber\label{rho_pol_qqbar}
\Pi_{V8}^{(all)} &=& 2.3 \times 10^{-7} M_B^{10} - 2.6 \times
10^{-8} M_B^8 + 9.1 \times 10^{-8} M_B^6 + 3.5 \times 10^{-7} M_B^4
- 4.9 \times 10^{-8} M_B^2\, ,
\\ \nonumber \Pi_{T8}^{(all)} &=& 1.4 \times 10^{-6} M_B^{10} - 1.7 \times 10^{-7}
M_B^8 + 1.2 \times 10^{-7} M_B^6 - 4.3 \times 10^{-9} M_B^4 + 4.9
\times 10^{-8} M_B^2\, .
\\
\end{eqnarray}
%
From these equations, we find that better convergence is achieved
for $T_8$ than for $V_8$ in the region $1 \lesssim M_B^2 \lesssim 2$
GeV$^2$. The pole contributions are significantly improved as shown
in Table 1.

\begin{figure}[hbt]
\begin{center}
\scalebox{0.9}{\includegraphics{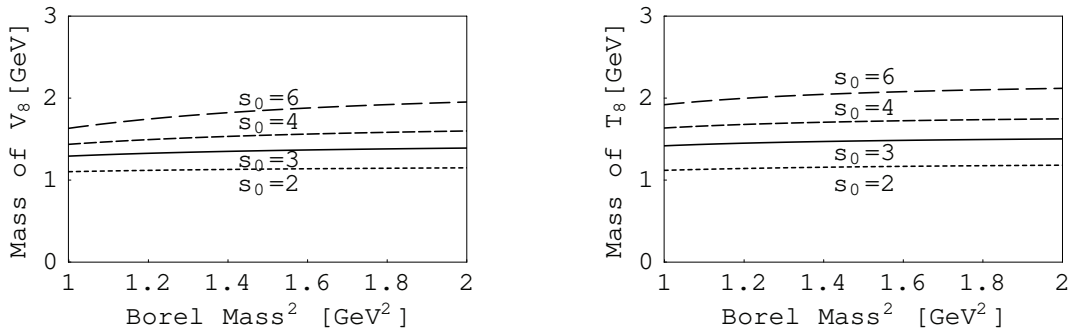}} \caption{Mass of
the tetraquark calculated by the currents $V_8$(Left) and
$T_8$(Right), as functions of the Borel mass square $M_B^2$, for
several threshold values $s_0 = 2, 3, 4$ and 6 GeV$^2$.}
\label{pic_mass_mesonic_single}
\end{center}
\end{figure}

In Fig.~\ref{pic_mass_mesonic_single}, we show the masses of the
tetraquark currents $V_8$ and $T_8$ as functions of the Borel mass
for several threshold values $s_0$ (Borel curves). As in the case of
$T_3$ current, the Borel stability seems good but the result depends
on the threshold value $s_0$. However, once again, if we take the
threshold value at $s_0 \sim 3$ GeV$^2$, the mass of the tetraquark
turns out to be reasonable, though the precise values are slightly
smaller: the mass of $T_8 \sim 1.5$ GeV and the mass of $V_8 \sim
1.4$ GeV.

%
%

%
\subsection{Analysis of Mixed
Currents}\label{sec_mixed_current}
%

In order to improve the sum rule, we attempt to make linear
combinations of independent currents for both diquark and mesonic
currents. Since linear combinations of five currents contain ten
mixing angles, the full consideration with these ten parameters is
rather cumbersome. Instead, we make a linear combination of two
currents $J_1$ and $J_2$ (any two from the independent currents),
$\eta = \cos \theta J_1 + \sin \theta J_2$, where $\theta$ is a
mixing angle. Then the correlation functions are written as
%
\begin{equation}
\bra \eta \eta^\dagger \ket =
\cos^2 \theta \bra J_1 J_1^\dagger \ket
+
\sin^2 \theta \bra J_2 J_2^\dagger \ket
+
\cos \theta \sin \theta \bra J_1 J_2^\dagger \ket
+
\cos \theta \sin \theta \bra J_2 J_1^\dagger \ket
\, .
\label{correlator_mixed}
\end{equation}
%
The mixing is chosen with the following requirements:
%
\begin{enumerate}

\item
The OPE has a good convergence as going to terms of higher
dimensional operators.

\item
The spectral density becomes positive for all (or almost all) $s$
values, and then $\Pi(M_B^2)$ becomes positive for all Borel mass
and threshold values.

\item
Pole contribution is sufficiently large.

\end{enumerate}
%

We have tried various combinations of two currents to realize good
sum rules. While doing so, we have realized that the diquark
currents are more independent than the mesonic currents. This means
that the cross terms of (\ref{correlator_mixed}) have only a
minor contribution for diquark currents, while they have a large
contribution for mesonic currents.

According to the requirement (1), we would like to make a linear
combination such that the highest dimensional (eight) term is
suppressed. For diquark currents, we find it convenient to take two
combinations:
%
\be \eta &=& \cos \theta A_3 + \sin \theta V_6 \, ,
\label{eta1} \\
\xi &=& \cos \theta P_6 + \sin \theta S_6 \, .
\label{xi}
\ee
%
By choosing $\cot \theta \sim \sqrt{2}$, we find that the term of
dimension eight of (\ref{eta1}) is suppressed, while for  $\cot
\theta \sim 1$, the term of dimension eight of (\ref{xi}) is
suppressed. The Borel transformed correlation function of (\ref{xi})
$\Pi_\xi(M_B^2)$, however, takes negative values. Therefore, this
current should be rejected for the sum rule analysis. In this way we
are led to the current $\eta$ of (\ref{eta1}). From now on,  we will
denote $\eta \to \eta_1$.

For the mesonic case, it turns out that the cross term contributions
are large. Accordingly, we attempt a complex angle to improve the
sum rule analysis. By choosing $t_1 = 0.91,~t_2 = -0.41$, we
construct a current:
\begin{eqnarray}
\eta_2 &=& S_1 + (t_1 +
\mathbf{i}t_2) P_1 \label{eta2} \, .
\end{eqnarray}
The numerical Borel transformed correlation functions are
%
\begin{eqnarray}\nonumber
\Pi_{1}^{(all)} &=& 1.1 \times 10^{-6} M_B^{10} - 1.3 \times 10^{-7}
M_B^8 + 4.8 \times 10^{-7} M_B^6 - 2.0 \times 10^{-8} M_B^4 + 5.2
\times 10^{-9} M_B^2\, ,
\\ \nonumber \Pi_{2}^{(all)} &=& 5.0 \times 10^{-7} M_B^{10} - 6.0 \times 10^{-8}
M_B^8 + 8.4 \times 10^{-8} M_B^6 - 2.2 \times 10^{-8} M_B^4 + 8.3
\times 10^{-9} M_B^2\, ,
\\
\end{eqnarray}
%
which may be compared with the previous results of
(\ref{rho_pol_qq}) and (\ref{rho_pol_qqbar}). It looks that the
convergence of the series is improved significantly. Therefore, we
can choose a smaller Borel mass square down to $M_B^2 \gtrsim 0.7$
GeV$^2$, where the pole contribution will be further increased up to
around 50 \%, and the convergence is still maintained.

%
\begin{figure}[hbt]
\begin{center}
\scalebox{1}{\includegraphics{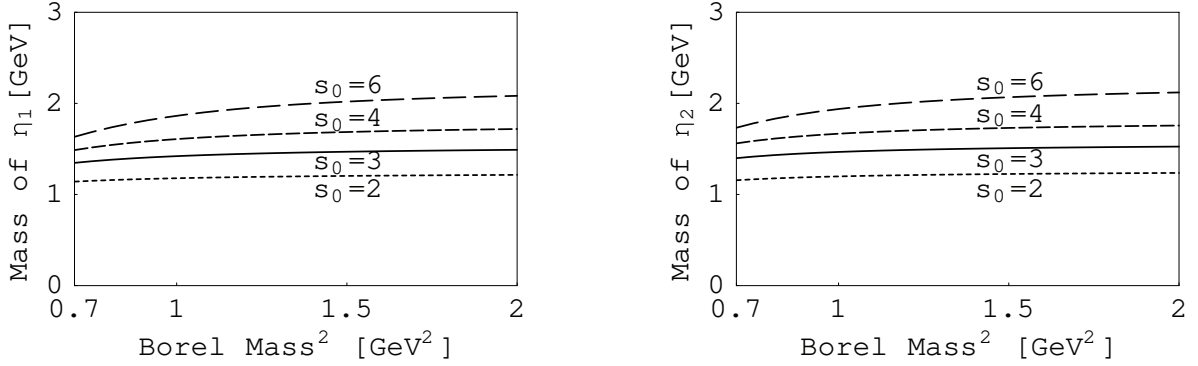}} \caption{Mass of the
tetraquark calculated by the mixed currents $\eta_1$(Left) and
$\eta_2$(Right), as functions of the Borel mass square $M_B^2$ for
several threshold values $s_0 = 2, 3, 4$ and 6 GeV$^2$.}
\label{pic_mass_mixing}
\end{center}
\end{figure}
%

\begin{figure}[hbt]
\begin{center}
\scalebox{1}{\includegraphics{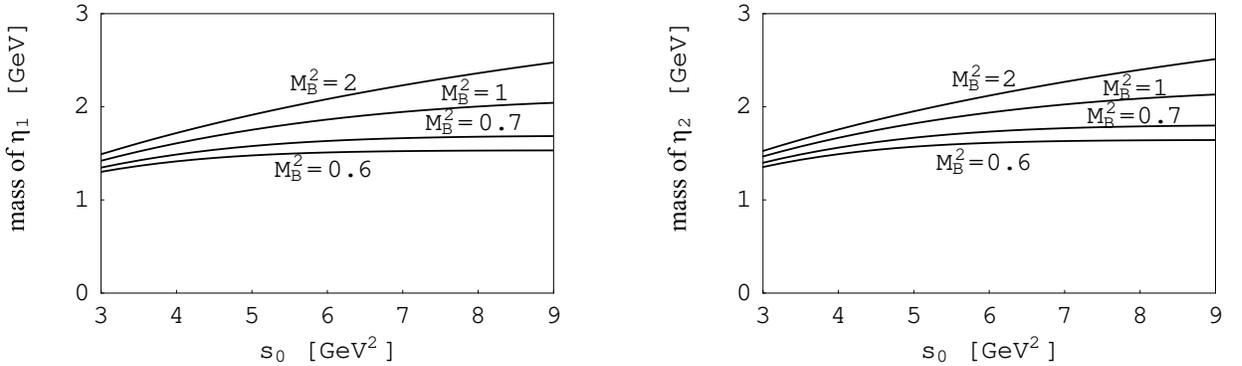}} \caption{Mass of the
tetraquark calculated by the mixed currents $\eta_1$(Left) and
$\eta_2$(Right), as functions of the threshold value $s_0$ for
several Borel mass square $M_B^2 = 0.6, 0.7, 1$ and 2 GeV$^2$.}
\label{pic_mass_s0}
\end{center}
\end{figure}

In Fig.~\ref{pic_mass_mixing}, we show the mass calculated from
$\eta_1$ and $\eta_2$ as functions of the Borel mass square for
several threshold values $s_0$. The Borel stability is improved from
the cases of the single currents. From these figures, we might think
that there is still a substantial $s_0$ dependence. However, this
dependence will be largely reduced if we choose a small Borel mass,
where the pole contribution is sufficiently large. In
Fig.~\ref{pic_mass_s0}, we show the mass calculated from $\eta_1$
and $\eta_2$ as functions of the threshold value for several Borel
masses. When $M_B^2 = 0.7$ GeV$^2$, the curve is very stable.
Moreover, the pole contribution is around 50 \%, and the convergence
is still maintained. Therefore, we obtain a very good sum rule,
where we find the mass calculated from the two currents $\eta_1$ and
$\eta_2$ is about 1.5 GeV. As the Borel mass increases, the pole
contribution decreases, and accordingly, the threshold dependence
becomes bigger.

%
%

\begin{figure}[hbt]
\begin{center}
\scalebox{0.6}{\includegraphics{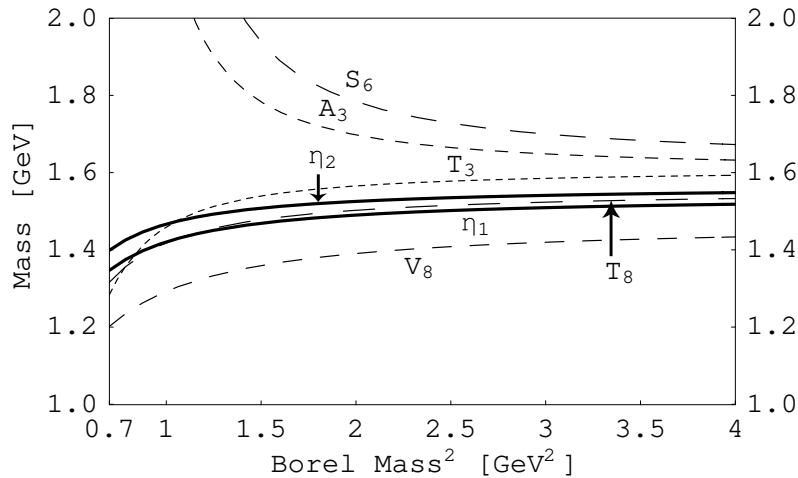}} \caption{Mass of the
tetraquark calculated by the currents $\eta_1$, $\eta_2$, $A_3$,
$S_6$, $T_3$, $V_8$ and $T_8$, as functions of the Borel mass square
$M_B^2$ in the region $0.7 < M_B^2 < 4$ GeV$^2$, for threshold value
$s_0 = 3$ GeV$^2$.} \label{pic_mass}
\end{center}
\end{figure}

Finally, in order to summarize our analysis, we show in
Fig.~\ref{pic_mass} masses of the tetraquark calculated by several
reasonable currents used in the present study as functions of the
Borel mass square at $s_0 = 3$ GeV$^2$. They are $S_6$, $A_3$ and
$T_3$ for the diquark construction, $T_8$ and $V_8$ for the mesonic
construction, and $\eta_1$ and $\eta_2$ for the mixed currents. The
plots are extended to a wider region of $M_B^2$ up to 4 GeV$^2$,
where the masses predicted by different currents tend to a same
value. We verify once again a good Borel mass stability for the
mixed currents, while some of the single currents show good
stability also ($T_3, T_8$ and $V_8$). The mass values varies
slightly, while we expect the mass of the tetraquark around 1.5 GeV.

%
%

%
\section{Summary}
%

We have presented a QCD sum rule study of the $ud\bar{s}\bar{s}$
tetraquark of $J^P=0^+$ and $I=1$, both in the diquark ($(\bar q
\bar q) (qq)$) and mesonic ($(\bar q q) (\bar qq)$) constructions.
We have found that in this channel of tetraquark, there are five
independent currents, which is shown both in the diquark and mesonic
constructions. For each single current, we have tested the sum rule
analysis, but it is found that not all of them provide a good
stability.

As an attempt to improve the stability of the sum rule, we have
considered linear combinations of independent currents. In order to
simplify the analysis, we took a superposition of various
combinations of two currents. Among them, we have found two cases
that lead to good sum rules, where we investigated $s_0$ (threshold
value) and $M_B$ (Borel mass) dependence, and convergence of OPE. A
good Borel stability is achieved in the region $0.7 \lsim M_B^2
\lsim 4$ GeV$^2$. In order to obtain a large enough pole
contribution (50 \%) and reduce the threshold value dependence, we
have to reduce the Borel mass. However, to maintain the convergence
of OPE, we can not reduce it too largely. When Borel mass square
$M_B^2$ is around 0.7 GeV$^2$, we get a very good QCD sum rule,
where the mass of the tetraquark turns out to be around 1.5 GeV.

Despite the seemingly good Borel mass stability, we think that we
should investigate the following points more carefully. For
instance, estimation of higher dimensional terms of ${\cal O}(1/s)$
could be important. Although we are able to construct the two mixed
currents such that the higher order contributions (in the present
calculation of OPE) of dimension six and eight terms are suppressed,
the question still remains concerning even higher order
contributions. Another question is the contribution of $KK$
scattering states, since the mass of the tetraquark is around $1.5$
GeV, and it can fall apart into the $KK$ states. Such a contribution
can be estimated by using the method proposed in
Refs.~\cite{Lee:2004xk,Kwon:2005fe}. These will be further
investigated in the future work.

\section*{Acknowledgments}
%
We thank Daisuke Jido for discussions on their work of
Ref.~\cite{Kojo:2006bh} and on general aspects of the QCD sum rule.
H.X.C is grateful to the Monkasho fellowship for supporting his stay
at Research Center for Nuclear Physics where this work is done. A.H.
is supported in part by the Grant for Scientific Research ((C)
No.16540252) from the Ministry of Education, Culture, Science and
Technology, Japan. S.L.Z. was supported by the National Natural
Science Foundation of China under Grants 10375003 and 10421003,
Ministry of Education of China, FANEDD, Key Grant Project of Chinese
Ministry of Education (NO 305001) and SRF for ROCS, SEM.
%
\appendix
%

%
\section{Five Independent Currents in $(\bar{q}q)(\bar{q}q)$ Basis}\label{app_five_currents}
%

We attempt to write a diquark current of
(\ref{define_diquark_current}) as a sum of $(\bar qq)$ mesonic pairs
($q = u, d, s)$,
%
\begin{eqnarray}
L_{\mu \nu \rho \sigma} \times \bar{s}_a^\mu \bar{s}_b^\nu u_c^\rho
d_d^\sigma = \sum_{n, i,j} C_i^1 C_j^2 (\bar{s}_b \lambda^n \Gamma_i
d_d) (\bar{s}_a \lambda^n \Gamma_j u_c) \, ,
\label{mesonic1}
\end{eqnarray}
%
where $\Gamma_i$ are the five Dirac matrices and $\lambda^n (n = 1,
\cdots ,8)$ are color matrices forming color singlet and octet
states out of $\mathbf{3} \times \mathbf{\bar 3}$. Therefore, in
(\ref{mesonic1}), the sum runs over ten terms of five $\Gamma_i$
matrices and two $\lambda^n$ combinations. They are
%
\begin{eqnarray}
\nonumber\label{define_meson_current} &&
S_1 =
(\bar{s}_au_a)(\bar{s}_bd_b)\, ,~~~~~~~~~~~~~~~~~~~~~~~~~~~~
S_8 =
(\bar{s}_a{\lambda^n_{ab}}u_b)(\bar{s}_c{\lambda^n_{cd}}d_d)\, ,
\\ \nonumber &&
V_1 = (\bar{s}_a\gamma_\mu u_a)(\bar{s}_b\gamma^\mu
d_b)\, ,~~~~~~~~~~~~~~~~~~~~~~
V_8 = (\bar{s}_a\gamma_\mu
{\lambda^n_{ab}} u_b)(\bar{s}_c\gamma^\mu {\lambda^n_{cd}} d_d)\, ,
\\ &&
T_1 =
(\bar{s}_a\sigma_{\mu\nu}u_a)(\bar{s}_b\sigma^{\mu\nu}d_b)\,
,~~~~~~~~~~~~~~~~~~~
T_8 = (\bar{s}_a\sigma_{\mu\nu} {\lambda^n_{ab}}
u_b)(\bar{s}_c\sigma^{\mu\nu} {\lambda^n_{cd}} d_d)\, ,
\\ \nonumber &&
A_1 = (\bar{s}_a\gamma_\mu\gamma_5u_a)(\bar{s}_b\gamma^\mu\gamma_5d_b
)\, ,~~~~~~~~~~~~~~~
A_8 = (\bar{s}_a\gamma_\mu\gamma_5
{\lambda^n_{ab}} u_b)(\bar{s}_c\gamma^\mu\gamma_5 {\lambda^n_{cd}}
d_d)\, ,
\\ \nonumber &&
P_1 =
(\bar{s}_a\gamma_5u_a)(\bar{s}_b\gamma_5d_b)\,
,~~~~~~~~~~~~~~~~~~~~~~
P_8 = (\bar{s}_a\gamma_5 {\lambda^n_{ab}}
u_b)(\bar{s}_c\gamma_5 {\lambda^n_{cd}} d_d) \, ,
\end{eqnarray}
%
where in the octet representation inner product of $\lambda^n$ ($n =
1, \cdots,  8$) is taken. The quark-antiquark pairs in different
currents have different properties:
%
\begin{eqnarray}
\nonumber && S_1: (J ^ P = 0 ^ +, 8_f, 1_c),~~~~~~~~~~~~~~~~~
S_8:(J
^ P = 0 ^ +, 8_f, 8_c),
\\ \nonumber &&
V_1:(J ^ P = 1 ^ -, 8_f,
1_c),~~~~~~~~~~~~~~~~~
V_8:(J ^ P = 1 ^ -, 8_f, 8_c),
\\ \nonumber &&
T_1:(J ^ P = 1 ^ + \& 1 ^ -, 8_f,
1_c),~~~~~~~~~~~
T_8:(J ^ P = 1 ^ + \& 1 ^ -, 8_f, 8_c),
\\ \nonumber &&
A_1:(J ^ P = 1 ^ +, 8_f,
1_c),~~~~~~~~~~~~~~~~
A_8:(J ^ P = 1 ^ +, 8_f, 8_c),
\\ \nonumber &&
P_1:(J ^ P = 0 ^ -, 8_f, 1_c),~~~~~~~~~~~~~~~~~
P_8:(J ^ P = 0 ^ -,
8_f, 8_c).
\end{eqnarray}
%

In order to establish the five independent currents, first we change
their color structures
%
\begin{eqnarray}\nonumber
&&(\bar{s}_a u_b)(\bar{s}_b d_a) = \frac{1}{3} (\bar{s}_a
u_a)(\bar{s}_b d_b) + \frac{1}{2} (\bar{s}_a u_b)(\bar{s}_c
d_d){\lambda_{ab}}{\lambda_{cd}}\, ,
\\ &&(\bar{s}_a u_d)(\bar{s}_c
d_b){\lambda_{ab}}{\lambda_{cd}} = \frac{16}{9} (\bar{s}_a
u_a)(\bar{s}_b d_b) - \frac{1}{3} (\bar{s}_a u_b)(\bar{s}_c
d_d){\lambda_{ab}}{\lambda_{cd}}\, .
\end{eqnarray}
%
Then we use the Fierz transformation~\cite{Maruhn:2000af}
%
\begin{eqnarray}\nonumber
&&\frac{1}{3} (\bar{s}_a u_a)(\bar{s}_b d_b) + \frac{1}{2}
(\bar{s}_a u_b)(\bar{s}_c d_d){\lambda_{ab}}{\lambda_{cd}}
\\ = && (\bar{s}_a u_b)(\bar{s}_b d_a) \\ \nonumber = && -
\frac{1}{4}\{(\bar{s}_au_a)(\bar{s}_bd_b) + (\bar{s}_a\gamma_\mu
u_a)(\bar{s}_b\gamma^\mu d_b) + \frac{1}{2}
(\bar{s}_a\sigma_{\mu\nu}u_a)(\bar{s}_b\sigma^{\mu\nu}d_b)
\\ \nonumber && -
(\bar{s}_a\gamma_\mu{}\gamma_5u_a)(\bar{s}_b\gamma^\mu{}\gamma_5d_b
) + (\bar{s}_a\gamma_5u_a)(\bar{s}_b\gamma_5d_b)\}\, .
\end{eqnarray}
%
We obtain 10 equations in all
%
\begin{eqnarray}\label{fierz equations}\nonumber
&& \frac{1}{3} S_1 + \frac{1}{2} S_8 = -\frac{1}{4} \{ S_1 + V_1 +
\frac{1}{2} T_1 - A_1 + P_1 \}\, ,
\\ \nonumber && \frac{16}{9} S_1 - \frac{1}{3}
S_8 = -\frac{1}{4} \{S_8 + V_8 + \frac{1}{2} T_8 - A_8 + P_8 \}\, ,
\\ \nonumber && \frac{1}{3} V_1 + \frac{1}{2} V_8 = -\frac{1}{4} \{
4S_1 -2 V_1 - 2A_1 - 4P_1 \}\, ,
\\ \nonumber && \frac{16}{9} V_1 - \frac{1}{3} V_8 =- \frac{1}{4} \{ 4S_8 -2 V_8 - 2A_8 - 4P_8 \}\, ,
\\ && \frac{1}{3} T_1 + \frac{1}{2} T_8 =- \frac{1}{4} \{ 12S_1 -2 T_1
+ 12P_1 \}\, ,
\\ \nonumber && \frac{16}{9} T_1 - \frac{1}{3} T_8 =- \frac{1}{4} \{ 12S_8 -2 T_8 + 12P_8 \}\, ,
\\ \nonumber && \frac{1}{3} A_1 + \frac{1}{2} A_8 =- \frac{1}{4} \{-
4S_1 -2 V_1 -2 A_1 + 4P_1 \}\, ,
\\ \nonumber && \frac{16}{9} A_1 - \frac{1}{3} A_8 =- \frac{1}{4} \{-
4S_8 -2 V_8 -2 A_8 + 4P_8 \}\, ,
\\ \nonumber && \frac{1}{3} P_1 + \frac{1}{2} P_8 =- \frac{1}{4} \{S_1
-V_1 +\frac{1}{2} T_1 + A_1 + P_1 \}\, ,
\\ \nonumber &&
\frac{16}{9} P_1 - \frac{1}{3} P_8 =- \frac{1}{4} \{S_8 -V_8
+\frac{1}{2} T_8 + A_8 + P_8 \}\, .
\end{eqnarray}
%
Solving these linear equations, we find that there are five
independent currents. In other words, the rank of the $10 \times 10$
coefficient matrix is five. Any five currents among (\ref{mesonic1})
are independent and can be expressed by the other five currents. For
instance, we have the relations as
%
\begin{eqnarray}\nonumber
&&S_8 =  -\frac{7}{6}S_1 - \frac{1}{2}V_1 - \frac{1}{4}T_1 +
\frac{1}{2}A_1 - \frac{1}{2}P_1\, ,
\\ \nonumber && V_8 = - 2 S_1 + \frac{1}{3}V_1 + A_1 + 2 P_1\, ,
\\ && T_8 = - 6 S_1 + \frac{1}{3}T_1 - 6 P_1\, ,
\\ \nonumber && A_8 =  2 S_1 +
V_1 + \frac{1}{3}A_1 - 2 P_1\, ,
\\ \nonumber && P_8 =-
\frac{1}{2}S_1 + \frac{1}{2}V_1 - \frac{1}{4}T_1 - \frac{1}{2}A_1 -
\frac{7}{6}P_1\, .
\end{eqnarray}
%
Finally, we establish the relations between the diquark currents and
the mesonic currents. For instance, we can verify the relations
%
\begin{eqnarray}
\nonumber\label{eq_transformation}
&& S_6 = -\frac{1}{4} S_1 - \frac{1}{4} V_1 + \frac{1}{8} T_1 - \frac{1}{4} A_1 - \frac{1}{4} P_1\, , \\
\nonumber &&
V_6 = S_1 - \frac{1}{2} V_1 + \frac{1}{2} A_1 - P_1\, ,
\\ && T_3 = 3 S_1 + \frac{1}{2} T_1 + 3 P_1\, ,
\\ \nonumber && A_3 = S_1 + \frac{1}{2} V_1 - \frac{1}{2} A_1 - P_1\, ,
\\ \nonumber && P_6 = -\frac{1}{4} S_1 + \frac{1}{4} V_1 + \frac{1}{8} T_1 + \frac{1}{4} A_1
- \frac{1}{4} P_1\, .
\end{eqnarray}
%


\def\Ref#1{[\ref{#1}]}
\def\Refs#1#2{[\ref{#1},\ref{#2}]}
\def\npb#1#2#3{{Nucl. Phys.\,}{\bf B{#1}},\,#2\,(#3)}
\def\npa#1#2#3{{Nucl. Phys.\,}{\bf A{#1}},\,#2\,(#3)}
\def\np#1#2#3{{Nucl. Phys.\,}{\bf{#1}},\,#2\,(#3)}
\def\plb#1#2#3{{Phys. Lett.\,}{\bf B{#1}},\,#2\,(#3)}
\def\prl#1#2#3{{Phys. Rev. Lett.\,}{\bf{#1}},\,#2\,(#3)}
\def\prd#1#2#3{{Phys. Rev.\,}{\bf D{#1}},\,#2\,(#3)}
\def\prc#1#2#3{{Phys. Rev.\,}{\bf C{#1}},\,#2\,(#3)}
\def\prb#1#2#3{{Phys. Rev.\,}{\bf B{#1}},\,#2\,(#3)}
\def\pr#1#2#3{{Phys. Rev.\,}{\bf{#1}},\,#2\,(#3)}
\def\ap#1#2#3{{Ann. Phys.\,}{\bf{#1}},\,#2\,(#3)}
\def\prep#1#2#3{{Phys. Reports\,}{\bf{#1}},\,#2\,(#3)}
\def\rmp#1#2#3{{Rev. Mod. Phys.\,}{\bf{#1}},\,#2\,(#3)}
\def\cmp#1#2#3{{Comm. Math. Phys.\,}{\bf{#1}},\,#2\,(#3)}
\def\ptp#1#2#3{{Prog. Theor. Phys.\,}{\bf{#1}},\,#2\,(#3)}
\def\ib#1#2#3{{\it ibid.\,}{\bf{#1}},\,#2\,(#3)}
\def\zsc#1#2#3{{Z. Phys. \,}{\bf C{#1}},\,#2\,(#3)}
\def\zsa#1#2#3{{Z. Phys. \,}{\bf A{#1}},\,#2\,(#3)}
\def\intj#1#2#3{{Int. J. Mod. Phys.\,}{\bf A{#1}},\,#2\,(#3)}
\def\sjnp#1#2#3{{Sov. J. Nucl. Phys.\,}{\bf #1},\,#2\,(#3)}
\def\pan#1#2#3{{Phys. Atom. Nucl.\,}{\bf #1},\,#2\,(#3)}
\def\app#1#2#3{{Acta. Phys. Pol.\,}{\bf #1},\,#2\,(#3)}
\def\jmp#1#2#3{{J. Math. Phys.\,}{\bf {#1}},\,#2\,(#3)}
\def\cp#1#2#3{{Coll. Phen.\,}{\bf {#1}},\,#2\,(#3)}
\def\epjc#1#2#3{{Eur. Phys. J.\,}{\bf C{#1}},\,#2\,(#3)}
\def\mpla#1#2#3{{Mod. Phys. Lett.\,}{\bf A{#1}},\,#2\,(#3)}
\def\etal{{\it et al.}}

%

%

\end{document}